\documentclass[aps,prl,twocolumn,superscriptaddress]{revtex4}
\usepackage{graphicx}
\usepackage{times}
\usepackage{amsmath}
\usepackage{textcomp}
\usepackage{epstopdf}
\usepackage{float}

\makeatletter
\renewenvironment{widetext@grid}{%
  \par\ignorespaces
  \setbox\widetext@top\vbox{%
   \vskip15\p@
   \hb@xt@\hsize{%
    \leaders\hrule\hfil
    \vrule\@height6\p@
   }%
   \vskip6\p@
  }%
  \setbox\widetext@bot\hb@xt@\hsize{%
    \vrule\@depth6\p@
    \leaders\hrule\hfil
  }%
  \onecolumngrid
  \let\set@footnotewidth\set@footnotewidth@ii
}{%
  \par
  \twocolumngrid\global\@ignoretrue
  \@endpetrue
}%
\makeatother

\begin{document}

\title{An addressable quantum dot qubit with fault-tolerant control fidelity}

\author{M. Veldhorst}
\affiliation{Centre for Quantum Computation and Communication Technology, School of Electrical Engineering and Telecommunications, The University of New South Wales, Sydney, NSW 2052, Australia}
\author{J.C.C. Hwang}
\affiliation{Centre for Quantum Computation and Communication Technology, School of Electrical Engineering and Telecommunications, The University of New South Wales, Sydney, NSW 2052, Australia}
\author{C.H. Yang}
\affiliation{Centre for Quantum Computation and Communication Technology, School of Electrical Engineering and Telecommunications, The University of New South Wales, Sydney, NSW 2052, Australia}
\author{A.W. Leenstra}
\affiliation{University of Twente, PO Box 217, 7500 AE Enschede, The Netherlands}
\author{B. de Ronde}
\affiliation{University of Twente, PO Box 217, 7500 AE Enschede, The Netherlands}
\author{J.P. Dehollain}
\affiliation{Centre for Quantum Computation and Communication Technology, School of Electrical Engineering and Telecommunications, The University of New South Wales, Sydney, NSW 2052, Australia}
\author{J.T. Muhonen}
\affiliation{Centre for Quantum Computation and Communication Technology, School of Electrical Engineering and Telecommunications, The University of New South Wales, Sydney, NSW 2052, Australia}
\author{F.E. Hudson}
\affiliation{Centre for Quantum Computation and Communication Technology, School of Electrical Engineering and Telecommunications, The University of New South Wales, Sydney, NSW 2052, Australia}
\author{K.M. Itoh}
\affiliation{School of Fundamental Science and Technology, Keio University, 3-14-1 Hiyoshi, Kohoku-ku, Yokohama 223-8522, Japan.}
\author{A. Morello}
\affiliation{Centre for Quantum Computation and Communication Technology, School of Electrical Engineering and Telecommunications, The University of New South Wales, Sydney, NSW 2052, Australia}
\author{A.S. Dzurak}
\affiliation{Centre for Quantum Computation and Communication Technology, School of Electrical Engineering and Telecommunications, The University of New South Wales, Sydney, NSW 2052, Australia}

\date{\today}

\begin{abstract}
Exciting progress towards spin-based quantum computing\cite{1, 2} has recently been made with qubits realized using nitrogen-vacancy (N-V) centers in diamond and phosphorus atoms in silicon\cite{3}, including the demonstration of long coherence times made possible by the presence of spin-free isotopes of carbon\cite{4} and silicon\cite{5}. However, despite promising single-atom nanotechnologies\cite{6}, there remain substantial challenges in coupling such qubits and addressing them individually. Conversely, lithographically defined quantum dots have an exchange coupling that can be precisely engineered1, but strong coupling to noise has severely limited their dephasing times and control fidelities. Here we combine the best aspects of both spin qubit schemes and demonstrate a gate-addressable quantum dot qubit in isotopically engineered silicon with a control fidelity of 99.6$\%$, obtained via Clifford based randomized benchmarking and consistent with that required for fault-tolerant quantum computing\cite{7,8}. This qubit has orders of magnitude improved coherence times compared with other quantum dot qubits, with $T_2^*$ = 120 \textmu s and $T_2$ = 28 ms. By gate-voltage tuning of the electron $g^*$-factor, we can Stark shift the electron spin resonance (ESR) frequency by more than 3000 times the 2.4 kHz ESR linewidth, providing a direct path to large-scale arrays of addressable high-fidelity qubits that are compatible with existing manufacturing technologies.
\end{abstract}

\maketitle

The seminal work by Loss and DiVincenzo\cite{1} to encode quantum information using the spin states of semiconductor quantum dots generated great excitement, as it fulfilled what were then understood to be the key criteria\cite{2} for quantum computation, and has already led to the realization of 2-qubit operations such as the $\sqrt{\textrm{SWAP}}$\cite{9,10} and CPHASE\cite{11}. However, the limited lifetime and the associated fidelity of the quantum state represent a significant hurdle for the semiconductor quantum dot qubits realized thus far. A dephasing time up to $T_2^*$ = 37 ns\cite{12}, improved to $T_2^*$ = 94 ns\cite{13} using nuclear spin bath control, has been recorded for quantum dot spin qubits in GaAs/AlGaAs. A longer $T_2^*$ = 360 ns has been achieved using Si/SiGe quantum dots\cite{14}. The main strategy to improve these times has involved applying pulse sequences developed for bulk magnetic resonance, and we can specify a $T_2$ according to the applied pulse sequence. Using a Hahn echo sequence the coherence time of GaAs-based qubits has been extended to $T_2^H$ = 440 ns\cite{12}, with $T_2^H$ = 30 \textmu s achieved via pulse optimization\cite{15}, while the use of a Carr-Purcell-Meiboom-Gill (CPMG) pulse sequence has enabled a $T_2^{CPMG}$ = 200 \textmu s\cite{15}. Here, by realizing a quantum dot qubit in isotopically enriched silicon ($^{28}$Si), we remove the dephasing effect of the nuclear spin bath present in these previous studies, and show that all of the above coherence times can be improved by orders of magnitude. These long coherence times, in particular the dephasing time $T_2^*$, lead to  low control error rates and the high fidelities that will be required for large-scale, fault tolerant quantum computing\cite{7,8}. 

  In contrast with quantum dots, electron spin qubits localized on atoms or defects have been realized in almost spin-free environments, showing coherence times approaching\cite{4} and even exceeding seconds\cite{5}. However, coupling multiple qubits and addressing them individually will be highly non-trivial for these systems, but remain as key requirements for scalable quantum computation. While a recent proposal suggests that addressability might be possible using atom clusters\cite{16}, we demonstrate here a qubit that can be addressed and tuned via a simple gate voltage.  Strong spin-orbit coupling in InAs has enabled qubits to be realized in double quantum dots with distinct electron $g$-factors\cite{17}, but in silicon the spin-orbit coupling is much smaller\cite{18}. Despite this, the highly tunable quantum dot presented here allows us to vary the internal electric field by as much as 3 MV/m, resulting in a Stark shift that can tune the electron spin resonance (ESR) frequency by $>$ 8 MHz. Also, the long $T_2^*$ available in isotopically enriched silicon results in a  narrow ESR linewidth of 2.4 kHz. Consequently, we can tune the qubit operation frequency by more than 3000 times the minimum linewidth. These results, together with the inherent scalability of gated quantum dot qubits, open the possibility for large-scale and gate-voltage addressable qubit systems. Such systems can utilize existing technology for the manufacture of metal-oxide-semiconductor field-effect-transistors (MOSFETs) that constitute today’s computer processors.
	
\begin{figure*} [t!]
	\centering 
		\includegraphics[width=0.95\textwidth]{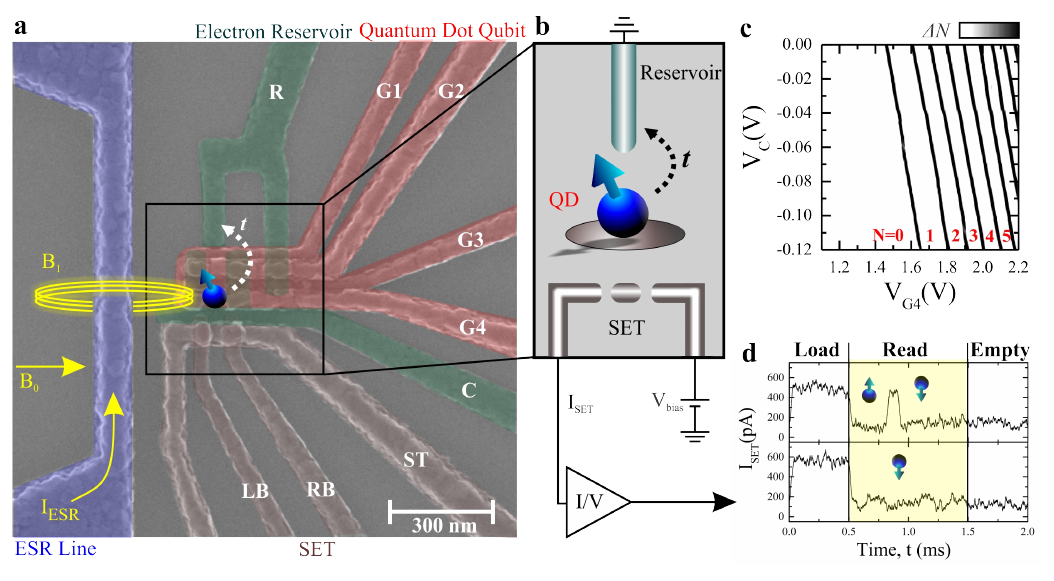}
		\caption{ \textbf{Silicon quantum dot qubit with SET readout and on-chip microwave spin control. a} SEM colored image and \textbf{b} schematic diagram of the device. The quantum dot structure (labels $C$ and $G$) can be operated as a single or double quantum dot by appropriate biasing of gate electrodes $G1$-$G4$.  The confinement gate $C$ runs underneath the gates $G2$-$G4$ and confines the quantum dot on all sides except on the reservoir side. Here, we operate the system in the single quantum dot mode, with the dot defined under $G4$ and tunnel coupled via $G3$ to the reservoir $R$. This provides maximum flexibility and the largest readout signal, as the dot is then closest to the SET (labels $ST$, $LB$ and $RB$). \textbf{c} Charge stability diagram. The SET is used as charge detector and a feedback loop is included to obtain maximum sensitivity. A square pulse of 40 mV peak-to-peak at 174 Hz is applied to $G4$. Grey-scale indicates the excess electron occupancy ($\delta N$) in the dot for each charge addition.  \textbf{d} By changing the voltage on $G4$, we can load and empty the quantum dot, performing spin readout in a single-shot measurement via energy-selective tunneling. All measurements were performed in a dilution refrigerator with base temperature $T \approx$ 50 mK and a dc magnetic field $B_0$ = 1.4 T.}  
		\label{fig:overview}
\end{figure*}

  Figure 1a and 1b show an SEM-image and the schematic design of the device, fabricated using a multi-level gate-stack silicon MOS technology\cite{19}. The device is fabricated on an epitaxially grown, isotopically purified $^{28}$Si epilayer with a residual concentration of $^{29}$Si at 800 ppm\cite{20}. We incorporated an on-chip transmission line\cite{21} to manipulate the spin-states of the dot using ESR pulses. The single electron transistor (SET) adjacent to the quantum dot structure is used as a sensor to monitor the electron occupancy within the quantum dot, as previously reported\cite{22}. The stability diagram for the quantum dot (Fig. 1c) is obtained by combining charge sensing with a feedback loop to keep maximum sensitivity and gate pulsing. The depletion of the last electron in the dot is observed in Fig. 1c, with no further charge transitions for $V_{G4}$ $<$ 1.6 V. In Fig. 1d, we show typical examples of single-shot spin readout measurements for the last electron, using spin-to-energy conversion\cite{23}. Further details of the spin readout measurements are provided in the Supplementary Information. All qubit spin statistics have been obtained via this method.
	
  To control and read the qubit state we make use of a two-level pulse sequence as shown in Fig. 2a. The tunnel coupling between dot and reservoir is tuned using the barrier gate $G3$ to yield a tunnel time $t \approx$ 100 \textmu s during the read phase.  There is almost no coupling in the control phase because the Zeeman-split spin states are plunged well below the Fermi level in the reservoir. We apply microwave pulses to the on-chip transmission line to create an ac magnetic field $B_1$ which drives transitions between the spin-down and spin-up states of the quantum dot\cite{24}. When B = 1.400 T, we find the resonance frequency $\nu_0 = (g^*\mu_B/h)B_{dc} \approx 39.1408$ GHz, resulting in $g^* = 1.998$. The qubit demonstrates coherent oscillations that coincide with $f_\uparrow =A \Omega^2/ \Omega_R^2 \sin^2(\Omega_R \tau/2)$, describing a qubit without decay and a visibility of $A$ = 0.7. Figure 2b shows sinusoidal Rabi oscillations obtained by varying the pulse length $\tau_p$ and Fig. 2c shows the oscillations while varying the frequency $\nu_{ESR}$. Confirmation that these are Rabi oscillations follows from the dependence $f_{Rabi} \propto B_1 \propto P_{ESR}^{1/2}$ (see inset Fig. 2c), where $P_{ESR}$ is the applied microwave source power and also from the increase in the Rabi frequency for non-zero detuning frequency (see Fig. 2d).

\begin{figure*} [t!]
	\centering 
		\includegraphics[width=0.95\textwidth]{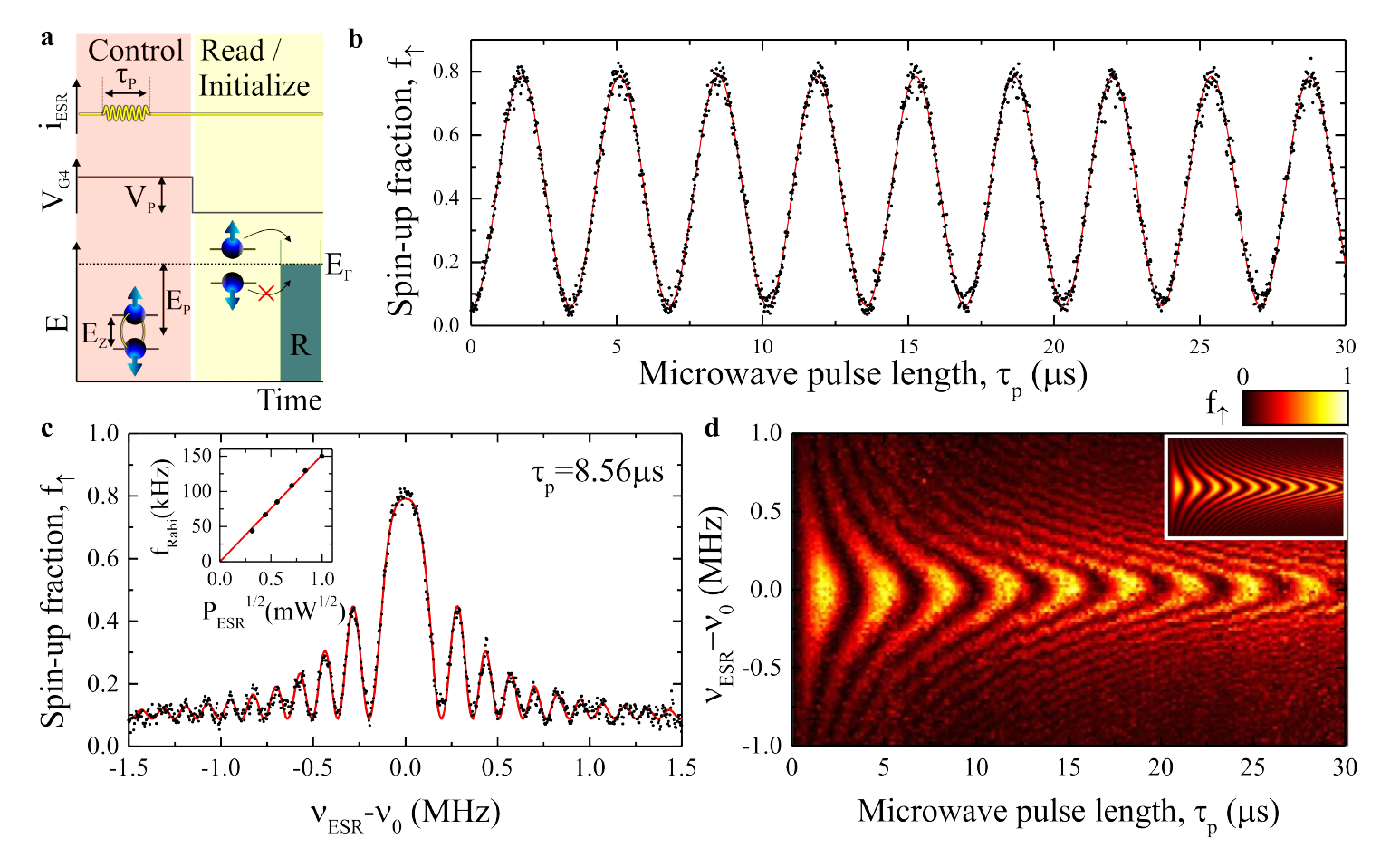}
		\caption{ \textbf{Electron spin resonance (ESR) and Rabi oscillations. a} Pulsing scheme for qubit control and readout. The Zeeman-split electron spin levels are plunged to an energy $E_P$ using gate $G4$ and ESR pulses are applied to rotate the spin on the Bloch sphere. Subsequently, the electron levels are increased to straddle the Fermi energy of the reservoir $R$, enabling spin readout. After readout the qubit is automatically initialized in the spin-down state. \textbf{b} Electron spin-up fraction $f_\uparrow$ as a function of the microwave burst duration $\tau_p$, with $P_{ESR}$ = 5 dBm. \textbf{c} Electron spin-up fraction $f_\uparrow$ as a function of ESR frequency around the resonance frequency, $\nu_0$ = 39.1408 GHz, with $\tau_p$ = 8.56 \textmu s (corresponding to the peak of the 3rd Rabi oscillation). \textbf{d} Color map of measured spin-up fraction $f_\uparrow$, showing Rabi oscillations as a function of $\tau_p$ for different microwave detuning frequencies. Inset: Corresponding calculated Rabi oscillations. All data in \textbf{b}-\textbf{d} is fitted by assuming no decay in time and using $f_\uparrow$ = $A \Omega^2/\Omega_R^2 \sin^2(\Omega_R \tau/2)$, with $\Omega$ and $\Omega_R$ the Rabi and total Rabi frequency, respectively. The visibility $A$ = 0.7, determined from the experimental data.}  
		\label{fig:overview}
\end{figure*}

	  When the detuning frequency is non-zero, coherent oscillations known as Ramsey fringes arise when the spin is pointing in the $x-y$ plane of the Bloch sphere. We detect these fringes by applying two $\pi$/2-pulses separated by a delay time $\tau$, followed by readout of the spin state. The resulting oscillations are shown in Fig. 3a and we extract a characteristic decay time $T_2^*$ = 120 \textmu s. The corresponding linewidth $1/\pi T_2^*$ = 2.6 kHz is close to the smallest measured ESR peak width $\delta \nu$ = (2.4 $\pm$ 0.2) kHz measured at $P_{ESR}$ = -20 dBm (see Fig. S2).  Slow environmental changes between individual single-shot readout events are one of the main factors leading to the decay of the Ramsey coherence fringes. To remove the effects of this noise we have applied a  Hahn-echo technique, where a $\pi_x$ pulse is applied exactly in between two $\pi_x/2$ pulses - see Fig. 3b. From this we measure a spin coherence time $T_2^H$ = 1.2 ms. The Hahn echo amplitude decays with an exponent $\eta$ = 2.2, indicating that the dominant source of decoherence is 1/$f$ noise. We can further increase the coherence time by applying a CPMG sequence, where a series of $\pi_y$ pulses are applied to refocus the signal. Figure 3c shows an echo decay obtained by applying 500 $\pi_y$ pulses, with a resulting coherence time of $T_2^{CPMG}$ = 28 ms.
  We now turn to the qubit fidelities (see the Supplementary Information for full details). The measurement fidelity $F_M$ = 92$\%$ and initialization fidelity $F_I$ = 95 $\%$ are primarily limited by broadening in the electron reservoir. We have characterized the control fidelity of the qubit via randomized benchmarking\cite{25} on Clifford gates, shown in Fig. 4. In this protocol, the fidelity of an individual Clifford gate is obtained by interleaving it with random Clifford gates and measuring the decay with increasing sequence length. The protocol ends with a final random recovery Clifford, such that the outcome is either spin up or spin down. A reference sequence without interleaved gates is performed to observe the decay due to the random Cliffords. By analyzing the data we find an average control fidelity of $F_C$ = 99.59$\%$, with all gates having an error rate below the 1$\%$ tolerance requirement for quantum error correction using surface codes\cite{8}. We note that the decay is slightly non-exponential, indicating dependent errors from a dephasing limited fidelity, which can possibly be removed by using composite and shaped pulse sequences, as routinely employed in NMR experiments. 
	
	\begin{figure*} [t!]
	\centering 
		\includegraphics[width=0.85\textwidth]{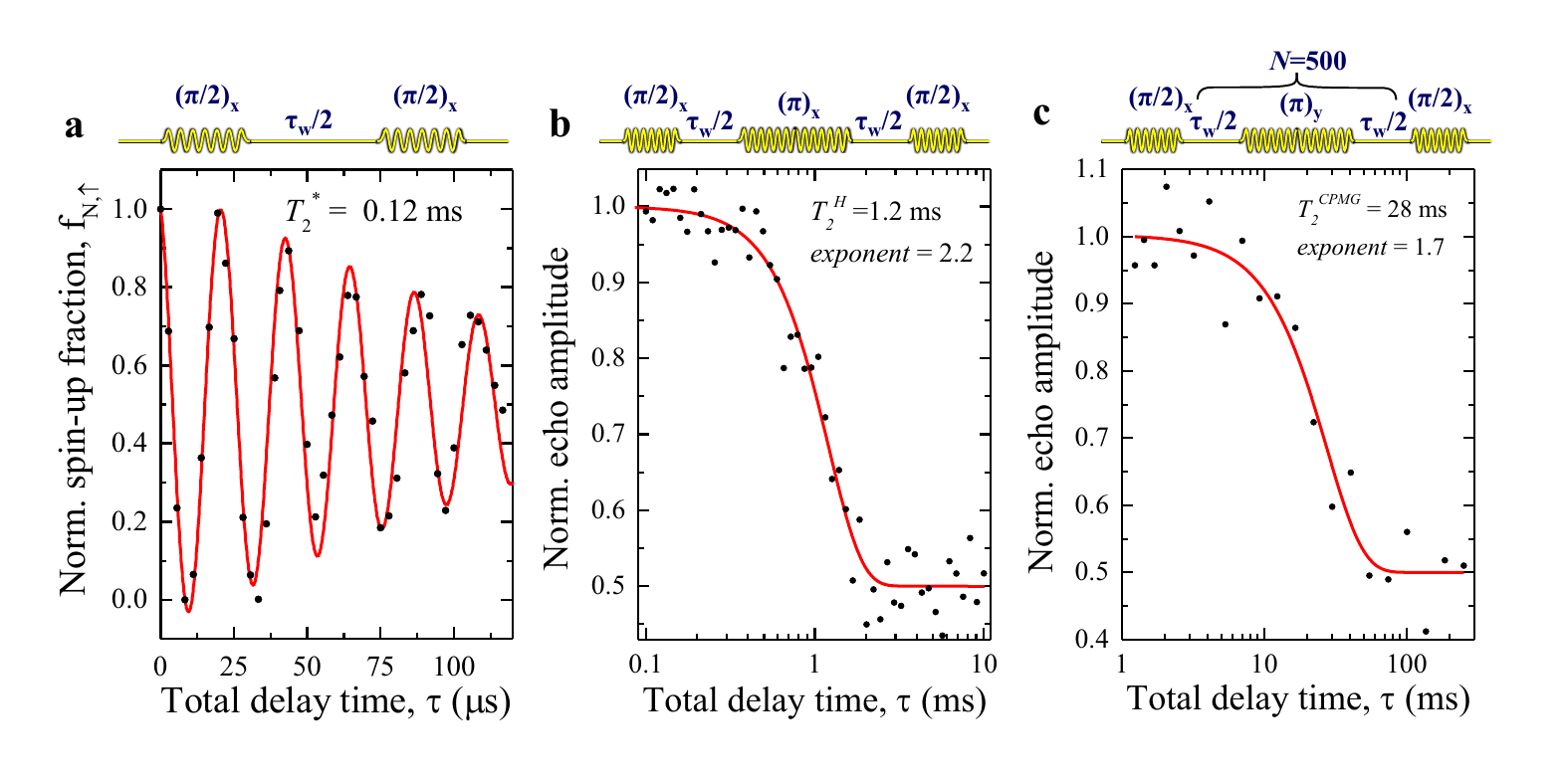}
		\caption{ \textbf{Qubit coherence. The spin state statistics are normalized with respect to the visibility to account for fluctuations between different measurements. a} By varying the delay time $\tau_w$ between two $\pi_x$/2 pulses (see inset), Ramsey oscillations arise in the spin-up fraction $f_\uparrow$. Fitting the decay with $f_{N,↑}=e^{-[\tau/T_2^* ]^\alpha}$, with $\alpha$ = 1.3, we deduce a dephasing time $T_2^*$ = 120 \textmu s. \textbf{b} A Hahn-echo pulse sequence incorporates an additional $\pi$-pulse (inset), and compensates for slow drifts in the environment. The resulting spin-up fraction $f_\uparrow$ decay gives the spin coherence time $T_2^H$ = 1.2 ms. \textbf{c} By applying a CPMG pulse sequence (inset) we can further enhance the coherence time, giving $T_2^{CPMG}$= 28 ms.}  
		\label{fig:overview}
\end{figure*}

  The vertical electric field $F_z$ in our quantum dot can be tuned over a large range by increasing the voltage on $G4$, while reducing the voltage on $C$ to maintain an electron occupancy of $N$ = 1. Recent experiments on silicon dots have observed an anticrossing of the spin and valley states (see inset to Fig. 5a) due to spin-orbit coupling, which occurs in a small energy window of neV to \textmu eV, depending on the interface roughness\cite{26,27}. Using a recently developed ‘hot-spot’ spin relaxation technique\cite{26} we have measured (Fig. S5) the magnitude of the valley splitting $E_{vs}$ as a function of gate voltage (Fig. 5a) and find a linear dependence of $E_{vs}$ upon $F_z$ that differs by only 12$\%$ from a device reported previously\cite{26}. 
	
  The same internal electric field that we use to tune the valley splitting can also be used to tune the qubit resonance frequency by more than 8 MHz (Fig. 5b, and Supplementary Information), corresponding to more than 3000 times the minimum observed ESR linewidth. This tunability, which is remarkable for a system with these long coherence times, provides encouraging prospects for scalability. We can operate our device in regimes both above and below the spin-valley anticrossing with no discernable impact on the ESR frequency dependence with $F_z$, demonstrating a gate-addressable and high-fidelity qubit well away from the valley anticrossing point, where the relaxation time dramatically drops\cite{26}.  The electric field creates a Stark shift of the electron $g^*$-factor due to the small but finite spin-orbit coupling. Tight binding simulations\cite{18} and measurements on donors in silicon\cite{28} indicate a quadratic Stark shift in $g^*$. By fitting our data we find a quadratic Stark coefficient of $\eta_2$ = 2.2 nm$^2$/V$^2$, comparable to that calculated in Rahman $et$ $al.$ \cite{18}. 
	
\begin{figure} [t!]
	\centering 
		\includegraphics[width=0.5\textwidth]{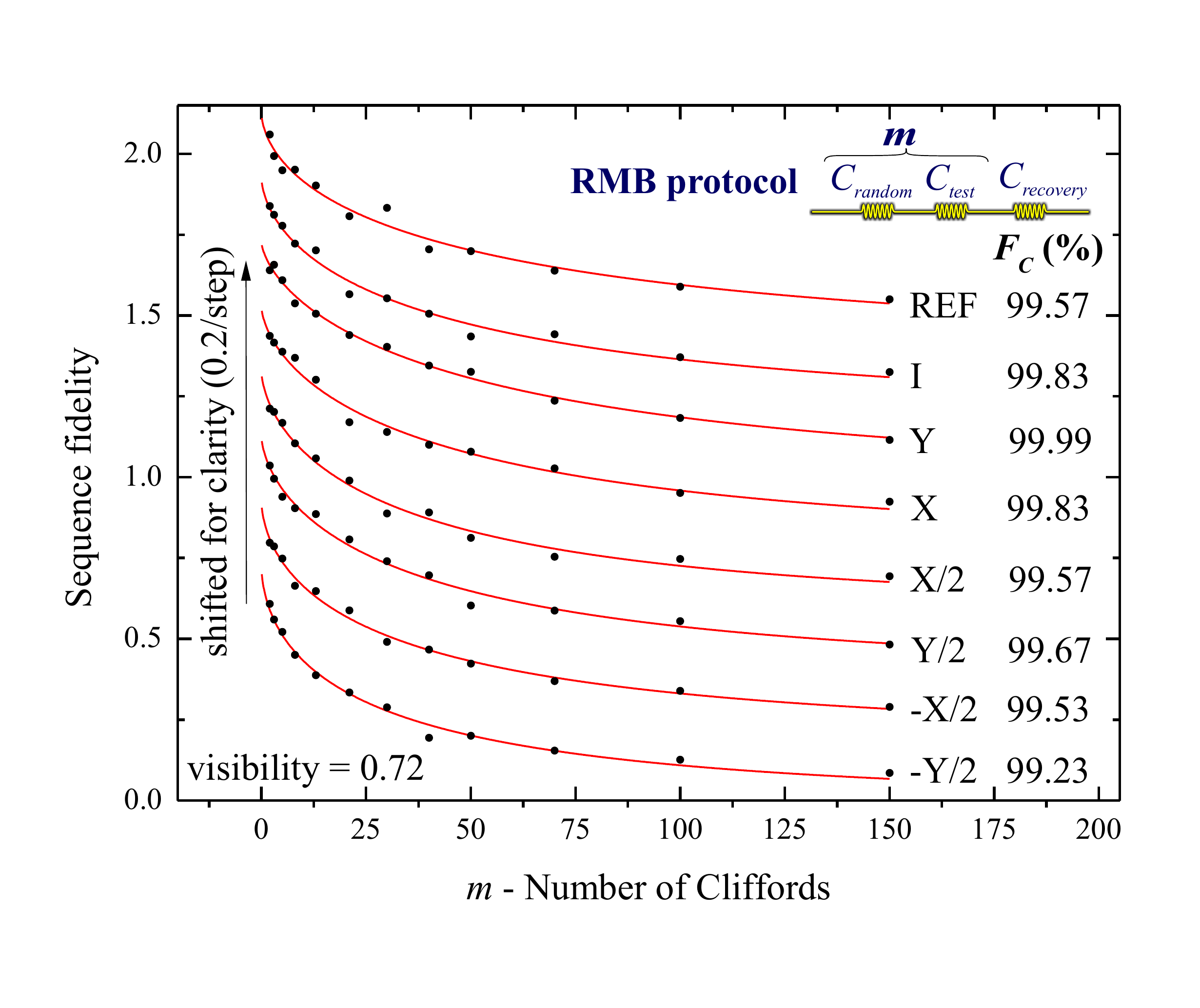}
		\caption{ \textbf{Control fidelity analysis via randomized benchmarking of Clifford gates.} The performance of each Clifford gate is tested by interleaving them with random Clifford gates. From the decay we infer an average fidelity of 99.59$\%$, above the threshold required for quantum error correction using surface codes\cite{8}. The sequence fidelity decays over more than 150 pulses, with m the number of Clifford gates applied. A $\pi$-pulse of 1.6 \textmu s and a waiting time of 500 ns between consecutive gates has been used. The data is vertically shifted by 0.2/step and the visibility of all data is $A$ = 0.72, limited by readout and initialization errors. Further details are provided in the Supplementary Information.}  
		\label{fig:overview}
\end{figure}

	\begin{figure} [t!]
	\centering 
		\includegraphics[width=0.5\textwidth]{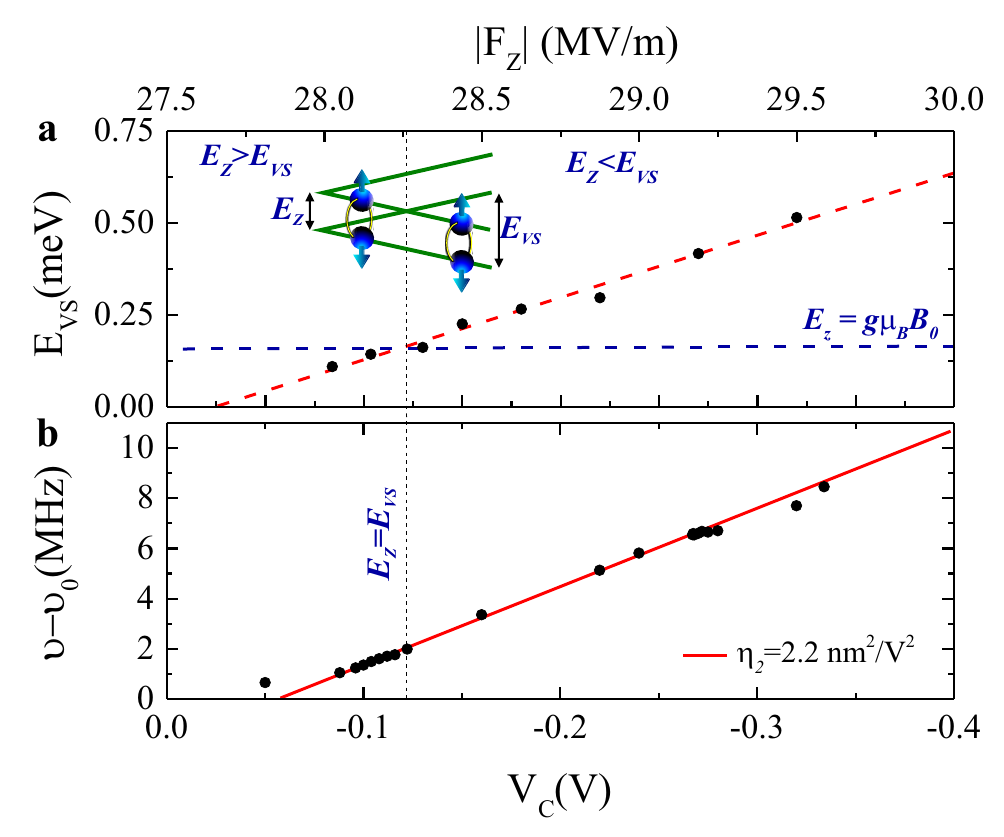}
		\caption{ \textbf{Gate-voltage tunability of the qubit operation frequency and of the valley splitting. a} Measured valley splitting $E_{vs}$, obtained using the method of Yang et al. \cite{26}, as a function of the confinement gate potential $V_C$. Inset: Electron spin states for the two-valley system, with an anticrossing when the Zeeman energy $E_z$ = $E_{vs}$. \textbf{b} Measured qubit resonance frequency as a function of $V_C$, due to a small but finite Stark shift. The red line is a fit using $g(|F_z|)/g(0) -1 = \eta_2|F_z|^2$, with $\eta_2$ = 2.2 nm$^2$/V$^2$. The vertical electric field $F_z$ (top axis) is calculated from the gate voltages using the numerical simulation of Yang $et$ $al.$ \cite{26}.}  
		\label{fig:overview}
\end{figure}

  The results presented here demonstrate that a single electron spin confined to a quantum dot in isotopically purified silicon can serve as a robust qubit platform for solid-state quantum computing. We have demonstrated long qubit coherence times, high fidelity control over the qubit, and the ability to individually address qubits via electrostatic gate-voltage control, meeting key criteria for quantum computation\cite{2}. The relevant coherence times ($T_2^*$, $T_2^H$ and $T_2^{CPMG}$) of our system exceed by two orders of magnitude the times of previous quantum dot qubits\cite{14,15}, while the fastest measured Rabi period of 400 ns combined with our $T_2^{CPMG}$ = 28 ms enables more than $10^5$ computational operations within the qubit coherence time. A recent experiment\cite{29} on a phosphorus donor qubit in $^{28}$Si found out that $T_2^{CPMG}$ is limited by Johnson-Nyquist thermal noise delivered via the on-chip ESR line, which is also the likely scenario for the quantum dot qubit given the comparable coherence times. This opens the possibility to increase coherence times even further. Faster qubit operations could be achieved by operating pairs of quantum dots as singlet-triplet qubits\cite{9}, with the potential to further increase the number of coherent operations. Such singlet-triplet qubits could not rely on a magnetic field gradient from lattice nuclear spins, since these are absent in isotopically enriched silicon, however a field gradient could be realized via an on-chip nanomagnet. The voltage-tunable Stark shift demonstrated here could also be exploited to create different effective $g^*$-factors for the individual dots. 
	
  Direct gate addressability opens the prospect for many qubits to be integrated on a single chip, with global ac magnetic fields applied via a cavity or on-chip transmission lines to realize single qubit operations. Two-qubit operations could then be achieved, for example, via gate-controlled exchange coupling between pairs of quantum dots. A recent model, applicable to our qubit system, predicts that 2-qubit gate operations with high fidelities are also possible\cite{30}. Taken together with the high control fidelities demonstrated for our 1-qubit gate, this now places quantum dot spin qubits as a viable candidate for fault-tolerant quantum computing. While scaling qubits involves complex wiring and will be a formidable task, one control line per qubit could be sufficient for the platform presented here. A confinement potential could be realized with one gate designed as a large grid, and one top gate for each qubit for addressing and controlling the exchange coupling to the other qubits. Finally, we note that the device structure used here can be modified to use poly-silicon gate electrodes and standard complementary metal-oxide-semiconductor (CMOS) manufacturing technologies that are currently used to fabricate more than one billion transistors on a single chip.
	
Acknowledgments: We thank D.J. Reilly and J.R. Petta for valuable discussions. We acknowledge support from the Australian Research Council (CE11E0096), the US Army Research Office (W911NF-13-1-0024), and the Australian National Fabrication Facility. M.V. acknowledges support from the Netherlands Organization for Scientiﬁc Research (NWO) through a Rubicon Grant.

Author contributions: M.V., J.C.C.H., C.H.Y, A.W.L., B.R., J.P.D. and J.T.M performed the experiments. M.V. and F.E.H. fabricated the devices. K.M.I. prepared and supplied the 28Si epilayer wafer. M.V., C.H.Y, A.M. and A.S.D. designed the experiments and analyzed the results. M.V. and A.S.D. wrote the manuscript, with input from all authors.

The authors declare no competing financial interests.

Correspondence should be addressed to

M.V. (M.Veldhorst@unsw.edu.au) or

A.S.D.(A.Dzurak@unsw.edu.au).

\clearpage
\begin{widetext}

\begin{center}
\textrm{\textbf{Supplementary Information}}

\textrm{\textbf{An addressable quantum dot qubit with fault-tolerant control fidelity}}

\end{center}

\end{widetext}

\textbf{Experimental methods}

The device is fabricated on an epitaxially grown, isotopically purified $^{28}$Si epilayer with a residual concentration of 800 ppm of $^{29}$Si\cite{20}. Using a multi-level gate-stack silicon MOS technology\cite{19}, three layers of Al-gates are fabricated with a thickness of 25, 50 and 80 nm, separated by thermally grown AlO.

The measurements were conducted in a dilution refrigerator with base temperature $T_{bath} \approx$ 50 mK. DC voltages were applied using battery-powered voltages sources and added through resistive voltage dividers/combiners to voltage pulses using an arbitrary waveform generator (LeCroy ArbStudio 1104). Filters were included for slow and fast lines (10Hz to 80MHz). ESR pulses were delivered by an Agilent E8257D microwave analog signal generator and a 3dBm attenuator at the 4K plate. The stability diagram is obtained using a double lock-in technique (Stanford Research Systems SR830) with dynamic voltage compensation\cite{22}.

All our qubit statistics are based on counting the spin states of the quantum dot, which is operated with an occupancy of one electron ($N$ = 1). Each data point represents the average of 500 up to 10000 single shot read outs, taken in 5 to 200 sweeps to compensate for small drifts.

\textbf{1. Device structure}

Figure 1a of the main text shows an SEM-image of the device. The device can be separated into four parts: the ESR line (purple), the SET (brown), the quantum dot structure (green and red) and the electron reservoir (red). The quantum dot structure is defined by gates $C$ and $G1$-$4$. Here, we operate the device in the single quantum dot regime, with the quantum dot formed under $G4$. The confinement gate $C$ surrounds gate $G4$. This allows us to change the gate voltage on one of the two gates and compensate with the other gate, thereby keeping the same electron occupancy while tuning the electric field (i.e. the size of the dot). In the single quantum dot mode, $G2$ and $G1$ are set to a high potential, such that a continuous 2-DEG is formed underneath these gates and the reservoir, $R$. The SET is used to measure charge transitions between the quantum dot and the reservoir and the ESR line is used to create an ac magnetic field to rotate the spin of the electron in the quantum dot. 

\textbf{2. Single-Shot Readout}

  The readout of the spin state of the quantum dot is performed using a spin to energy conversion\cite{23} method. In this scheme, the readout is performed with the spin-up level above the Fermi energy $E_F$ of the reservoir, and the spin-down level below $E_F$. At this bias position, only a spin-up electron can tunnel out from the dot to the reservoir, followed by the tunneling into the dot of a spin-down electron. In Fig. S1 we show a 2D-map of the SET readout signal $I_{SET}$ obtained by applying a 3-level pulse sequence (load, read and empty) to gate $G4$. The white ‘tail’, $\approx$100 \textmu s after the load pulse, corresponds to spin-dependent tunneling. Individual traces of the readout signal $I_{SET}$ are shown in Fig 1d of the main text. For the qubit control we use a 2-level pulse sequence on $G4$, consisting of the load and read phase. At the end of the read phase, the electron will be in the spin-down state, which we use as initialization of our qubit. 
 
	\begin{figure} 
	\centering 
		\includegraphics[width=0.5\textwidth]{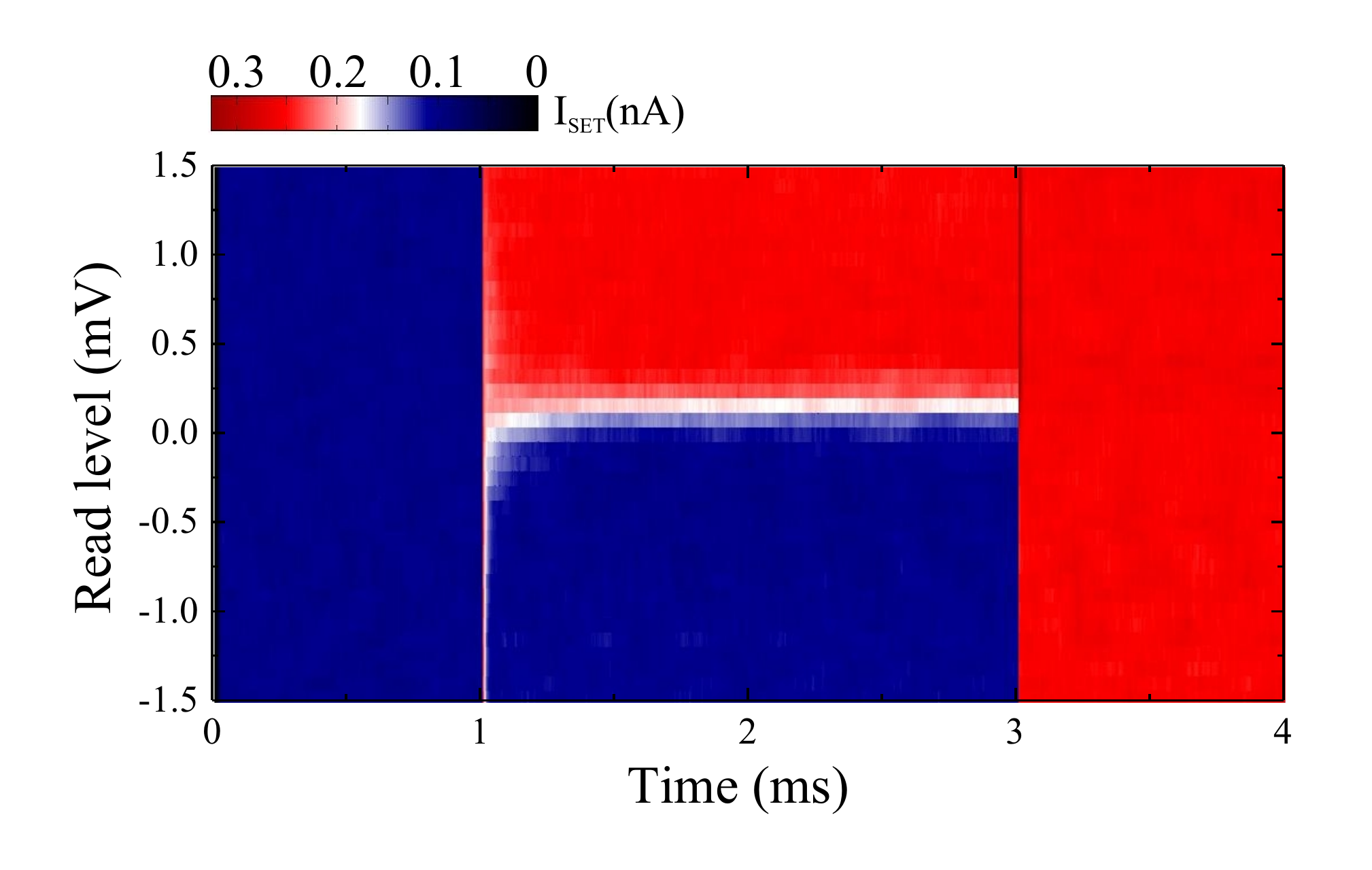}
		\caption{ \textbf{Single-shot readout of the last electron.} 2D color map showing the SET readout signal $I_{SET}$ when a 3-pulse sequence applied to $G4$, consisting of a load, read and empty phase. The read phase level is tuned from -1.5 to 1.5 mV around the Fermi energy of the reservoir $R$.  At 0 mV the Zeeman spin-split levels straddle the Fermi energy, and only a spin-up electron can tunnel out from the dot, followed by the tunneling in of a spin-down electron. The white ‘tail’ in the middle of the data map corresponds to the presence of single-shot spin readout events.}  
		\label{fig:overview}
\end{figure}

	\begin{figure} 
	\centering 
		\includegraphics[width=0.5\textwidth]{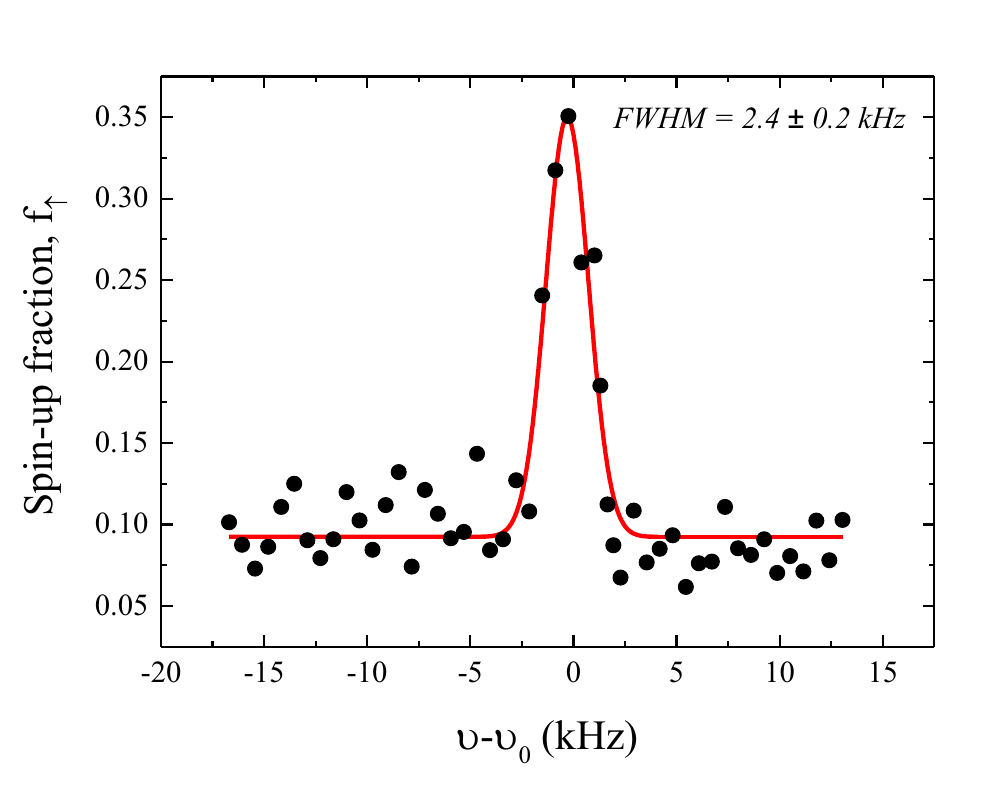}
		\caption{ \textbf{Minimum ESR linewidth.} At -20 dBm we find the narrowest ESR linewidth of (2.4 $\pm$ 0.2) kHz. This ESR linewidth is by far the narrowest obtained line width in any realized quantum dot qubit thus far and is the direct result of the long dephasing time in purified $^{28}$Si.}  
		\label{fig:overview}
\end{figure}

\textbf{3. Minimum ESR Linewidth}

  At high ESR powers, the ESR linewidth is strongly power broadened. By lowering the ESR power (to -20 dBm) we have been able to narrow the ESR linewidth down to $\delta \nu$ = (2.4 $\pm$ 0.2) kHz, as shown in Fig. S2. This corresponds to a $T_2^*$ = $1/\pi \delta \nu$ = (130 $\pm$ 10) \textmu s, close to the value of $T_2^*$ = 120 \textmu s obtained in the Ramsey experiment. This ESR linewidth is by far the narrowest obtained line width in any realized quantum dot qubit, and is the direct result of the long dephasing time in purified $^{28}$Si.

\textbf{4. Qubit Fidelity Analysis}

  The fidelity of our qubit can be divided into a measurement, initialization, and control fidelity. We have calculated the measurement and initialization fidelity according to the model of Pla $et$ $al.$ \cite{25}. We have obtained the control fidelity using randomized benchmarking. Here we explain briefly the results.

\textit{Measurement, initialization and intrinsic control fidelity}

 Our analysis is based on the Rabi measurement shown in Fig. 2c of the main text. Figure S3a shows a histogram of the spin readout as a function of the peak current after assigning a spin state to a single-shot readout trace (black dotted data). We have modeled these results using a simulation that includes readout errors due to white noise and thermal broadening \cite{25} and a small finite acoustic noise that is present in the measurement setup. The blue dashed, green dashed and red solid correspond to the respective simulations of the spin-up, spin-down and total spin states. From this simulation, we can extract the spin-down, spin-up and total error as a function of the current threshold, shown in Fig. S3b. The lowest total error we find is $\gamma$ = 0.127, where $\gamma_\downarrow$ = 0.106. These values, together with the probability $\alpha$ = 0.037 of a spin-down electron tunneling to the reservoir, thus creating a read out error, results in a measurement fidelity $F_M$ = $1-(\gamma+ \alpha(1- \gamma_\downarrow))/2 = 92 \%$. From the simulation of the spin readout histogram, we can extract the probability of erroneously initializing with a spin-up electron given by $\beta$ = 0.053. This results in an initialization fidelity of $F_I$ = 95 $\%$. The initialization error is created by a small random telegraph signal (RTS).

		\begin{figure}
	\centering 
		\includegraphics[width=0.5\textwidth]{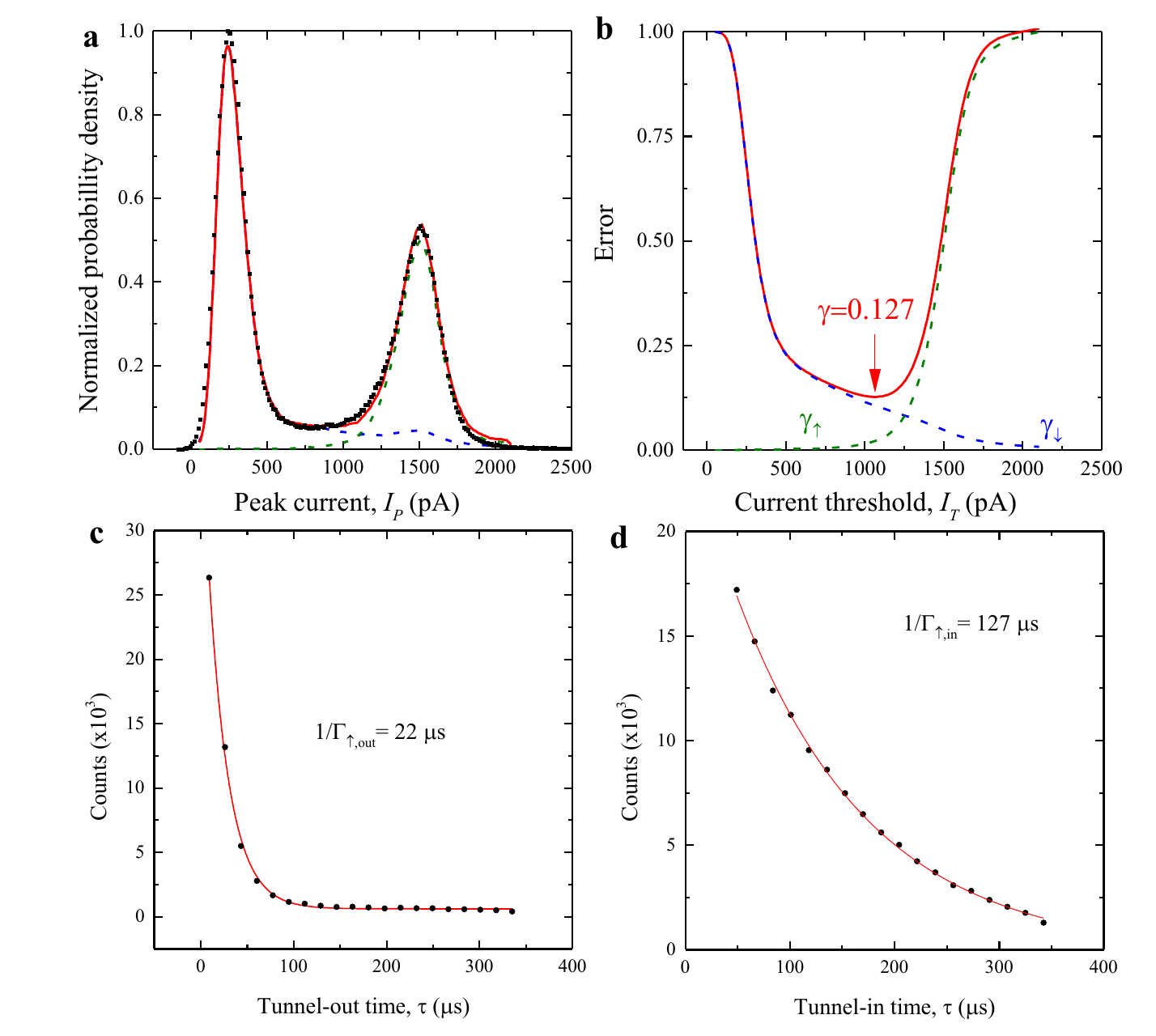}
		\caption{ \textbf{Peak histogram and tunnel times of single-shot readout.} We have used the same data as for the Rabi oscillations shown in Fig. 2c, to calculate the qubit fidelities. \textbf{a} Peak histogram as a function of the current. The black dotted line corresponds to the measurement, and the blue short-dashed, green dashed, and red solid line are simulations of the spin-down, spin-up, and total spin statistics, respectively. \textbf{b} From the simulation we can infer the spin-down, spin-up and total error. \textbf{c} Tunnel-out time of the electron from the quantum dot to the reservoir. \textbf{d} Tunnel-in time of an electron from the reservoir to the quantum dot.}  
		\label{fig:overview}
\end{figure}

\textit{Control fidelity via randomized benchmarking}

 We have investigated the qubit control by measuring the spin-up fraction decay as a function of the number of $\pi$-pulses in a CP sequence, shown in Fig. S4a. A single $\pi$-rotation corresponds here to 2.2 \textmu s and the waiting time $\tau_w$ = 500 ns, such that the decay due to $T_2$ (=28 ms) is negligible throughout the experiment. The decay observed over more than 400 rotations is fitted with $f_\uparrow=A+B e^{(-(N/N^*)\alpha)}$, where we have left $\alpha$ as a fitting parameter ($\alpha$ is found to be 1.3). From the fit we extract a characteristic decay $N^*$ = 140 oscillations, longer than a phosphorus electron qubit in $^{28}$Si and comparable to the results of the associated nuclear spin qubit\cite{29}. In the inset of Fig. S4a we also show the spin-up fraction after applying two $\pi$/2 pulses with increasing phase difference, demonstrating that the full Bloch sphere can be assessed with high fidelity.

To obtain the control fidelity we have employed a randomized benchmarking technique\cite{25} based on Clifford gates[31], shown in Fig. 4 of the main text and Fig. S4b. Each data point in the figure comprises 50 single shot readout events with 50 sweeps to average and 10 randomization sequences, corresponding to a total of 25000 records. The last Clifford gate of the randomization sequences ensures an outcome of either spin up or spin down. In this experiment, a $\pi$-rotation corresponds to 1.6 \textmu s and a waiting time of 500 ns between pulses has been used.  The sequence fidelity is obtained by taking the difference between the measured spin up and down states, such that any offset in the readout (due to initialization errors) is removed and the sequence fidelity at sequence length $m$ = 0 is the visibility $A$ = 0.72. We have fitted the data using $f_\uparrow=A e^{-(b m)^\alpha}$, with $b$ the sequence decay. We find $\alpha$ = 0.61, indicating time dependent errors[31,32]. This statement is reinforced by the observation that the $\pi$-pulses have higher fidelity then the $\pi$/2-pulses. This indicates that the fidelity can be further improved using composite and shaped pulses and decreasing the waiting time between pulses. As depolarization occurs half the time and the reference fidelity has average single qubit length 1.875, $F_{ref}=1-b/(2\times1.875) = 99.57\%$. The individual gate fidelities are obtained after subtracting the reference fidelity. The average gate fidelity is obtained by averaging all 24 Clifford fidelities and is $F = 99.59\%$. All fidelities have a standard deviation below 0.1.

\textbf{5. Valley Anticrossing Hot-Spot Measurements}

  The finite spin-orbit coupling present in silicon causes an anticrossing of the two $\Gamma$ valleys \cite{26}. Recently, Yang et al. \cite{26} demonstrated that the valley splitting energy can be controlled via the gate-tuned vertical electric field. Thereby, the magnetic field at which the two valleys have their anticrossing can be controlled using the gate voltages $C$ and $G4$. This valley anticrossing is important, as it mixes the spin states. Since spin is no longer a good quantum number close to the anticrossing point, the relaxation rate drastically increases in this regime, where it is only limited by the pure valley relaxation rate. This can be modeled in a two level system with the spin-orbit induced anticrossing gap $\delta_a$ at the off-diagonal axis, resulting in the following relaxation rates \cite{26}:
	
(i) for magnetic fields where the Zeeman energy is smaller than the valley splitting ($E_Z$ \textless $E_{VS}$):

\begin{flushleft}
$\Gamma_{\bar{2}1}=(\sqrt{\delta^2+\delta_a^2}-\delta)/(2\sqrt{\delta^2+\delta_a^2}) \Gamma_{v',v} ;$

\end{flushleft}

(ii) and for magnetic fields where $E_Z$ \textgreater $E_{VS}$:

\begin{flushleft}
$\Gamma=\Gamma_{\bar{3} 1}+\Gamma_{\bar{32}}=(\sqrt{\delta^2+\delta_a^2}+\delta)/(2\sqrt{\delta^2+\delta_a^2}) \Gamma_{v',v} + (\sqrt{\delta^2+\delta_a^2}-\delta)/(2\sqrt{\delta^2+\delta_a^2})  (\sqrt{\delta^2+\delta_a^2}+\delta)/(2\sqrt{\delta^2+\delta_a^2}) \Gamma_{v',v}$.

\end{flushleft}

	\begin{figure} 
	\centering 
		\includegraphics[width=0.5\textwidth]{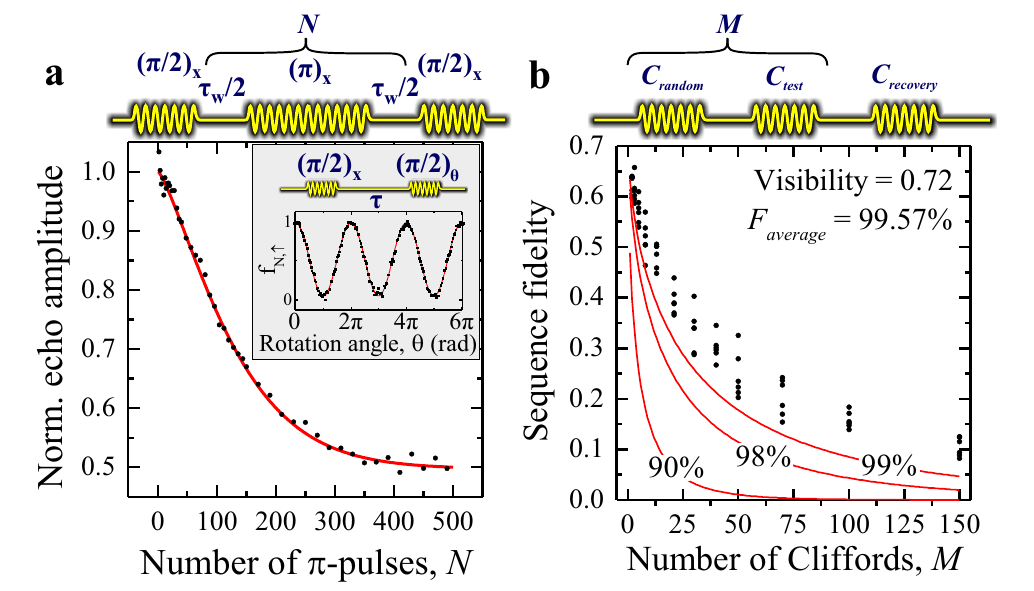}
		\caption{ \textbf{Clifford based randomized benchmarking and CP sequences. a} The spin up fraction (normalized with respect to the visibility) is measured as a function of number of $\pi$-pulses in a CP sequence. Inset: two $\pi$/2 pulses are applied with increasing phase difference, demonstrating that the full Bloch sphere can be assessed with high fidelity. Scatter in the data is due to read out errors, but the frequency can be accurately determined. \textbf{b} Clifford based randomized benchmarking. The data is the same as in the main text, but now plotted without shift and compared to different fidelity lines. The fidelity lines with corresponding fidelity $F$ are defined as $f_\uparrow=0.72 e^{-(2(1-F)m)^{0.62}}$ and subtracting the measured reference fidelity.}  
		\label{fig:overview}
\end{figure}

	\begin{figure} 
	\centering 
		\includegraphics[width=0.5\textwidth]{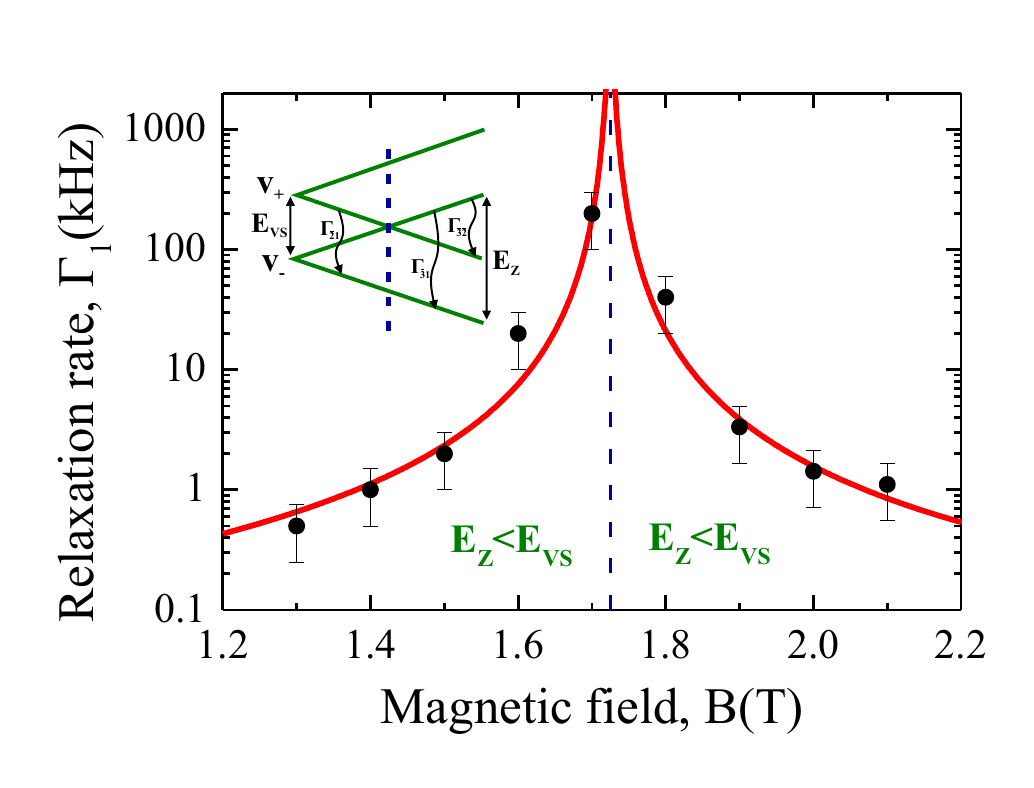}
		\caption{ \textbf{Hot-spot relaxation rate $\Gamma_1$ measurements to determine the valley anticrossing point.} By measuring the magnetic field dependence of the relaxation rate, we can determine the valley splitting energy $E_{VS}$. Using gate voltages $C$ and $G4$ we can tune the valley splitting to be larger or smaller than the Zeeman energy, as shown in Fig. 4a of the main text. From the fitting we infer a valley splitting energy with $E_{VS} /g\mu_B$ = 1.72 T, an anticrossing gap $\delta_a$ = 3$\times$10$^{-2}$ $E_{VS}$, and a pure valley relaxation rate $\Gamma_{v',v}$ = 2$\times$10$^7$ s$^{-1}$, in agreement with recent experiments\cite{26,27}. }  
		\label{fig:overview}
\end{figure}

Here, $\delta$ = ($E_{VS}$ - $E_Z$), and $\Gamma_{v',v}$ is the valley relaxation rate corresponding to the relevant transition. For large field variations, $\Gamma_{v',v}$ results in B$^5$ and B$^7$ dependencies, but close to the hot-spot, the spin-valley mixing terms dominate the relaxation rate dependence. Figure S5 shows a relaxation rate experiment to determine the hot-spot, showing a valley splitting energy $E_{VS}/g\mu_B$ = 1.72 T, an anticrossing gap $\delta_a$ = 3 $\times$ 10$^{-2}$ $E_{VS}$, and a pure valley relaxation rate $\Gamma_{v',v}$ = 2 $\times$ 10$^7$ s$^{-1}$, in agreement with recent experiments\cite{26,27}. We have found that $T_1$ is also dependent on both the plunge level and the exact voltages of the other gates, but we have verified that in all our experiments $T_1$ is the longest timescale.

\textbf{6. Tunability of the ESR Resonance Frequency}

In Fig. 5B of the main text we show the ESR resonance frequency dependence on the confinement gate $C$, while compensating $G4$ in order to keep an electron occupancy of $N$ = 1. We find that the resonance frequency can be tuned over 8 MHz, linearly varying with $C$, which is proportional to the electric field. We note that $G4$ and $C$ are approximately linearly related in this regime as well, but we have adjusted the other gates slightly to improve the readout during the measurements, so that $C$ represents the electric field with the highest accuracy. In principle, coupling to a nearby trap or a nuclear spin could cause a shift in the ESR resonance frequency via exchange ($J$) coupling and hyperfine ($A$) coupling, respectively. We rule out coupling to the reservoir, as changing the tunnel coupling over an order of magnitude (determined in the read phase) has no observable impact and the coupling should be well below 10 kHz during the control phase. From the stability diagram we observe a low level of disorder in the quantum dot. Furthermore, if the resonance frequency shift would be caused by coupling to a trap or a nuclear spin, we would expect to observe a shift when there is a spin-flip (and tunneling in the trap scenario). During the measurements (taken over months) no ESR resonance frequency shift (which should be larger than 8 MHz) is observed. We thereby conclude that coupling to a nearby spin is highly unlikely to be the cause of the observed ESR resonance frequency shift. This interpretation is reinforced by the fact that there is strong linear relationship with $C$ over a large range. The extracted Stark shift of $\eta_2$ = 2.2 nm$^2$/V$^2$ is furthermore comparable to tight binding simulations\cite{18} and bulk measurements on donors in silicon\cite{28}.

\end{document}